\pdfoutput=1
\documentclass[preprint,authoryear,12pt]{elsarticle}
\pdfoutput=1

\usepackage{epsfig}
\usepackage{aas_macros}
\usepackage{amssymb}
\usepackage{rotating}

\journal{Planetary \& Space Science}

\begin{document}

\begin{frontmatter}



\title{Astrophysical objects observed by the MESSENGER X-ray spectrometer during Mercury flybys}


\author{N P Bannister}\ead{nb101@le.ac.uk}
\author{G W Fraser}
\author{S T Lindsay}
\author{A Martindale}
\author{D L Talboys}

\address{Space Research Centre, Department of Physics \& Astronomy, University of Leicester, University Road, Leicester LE1 7RH, UK}

\begin{abstract}
The {\it MESSENGER} spacecraft conducted its first flyby of Mercury on 14th January 2008, followed by two subsequent encounters on 6th October 2008 and 29th September 2009, prior to Mercury orbit insertion on 18th March 2011. We have reviewed {\it MESSENGER} flight telemetry and X-ray Spectrometer observations from the first two encounters, and correlate several prominent features in the data with the presence of astrophysical X-ray sources in the instrument field of view. We find that two X-ray peaks attributed in earlier work to the detection of suprathermal electrons from the Mercury magnetosphere, are likely to contain a significant number of events that are of astrophysical origin. The intensities of these two peaks cannot be explained entirely on the basis of astrophysical sources, and we support the previous suprathermal explanation but suggest that the electron fluxes derived in those studies be revised to correct for a significant astrophysical signal. 

\end{abstract}


\end{frontmatter}


\section{Introduction}
On March 18th 2011 the MErcury Surface, Space ENvironment, GEochemistry, and Ranging ({\it MESSENGER}) spacecraft became the first probe to orbit the planet Mercury. Launched on August 3rd 2004, {\it MESSENGER}'s trajectory included six gravity-assist manoeuvres: one flyby of Earth (August 2nd 2005), two of Venus (October 24th 2006 and June 5th 2007), followed by three flybys of Mercury, on January 14th and October 6th 2008, and September 29th 2009. A detailed description of the mission is given by~\citet{Sol:2001,Sol:2007}. A review of the first Mercury encounter is given by~\citet{Sol:2008}, and a brief summary of the emerging view of Mercury after the three flybys is presented by~\citet{Sol:2010}.

The {\it MESSENGER} payload includes an X-ray spectrometer (XRS) for measurement of the planet's surface elemental composition.~\citet{Sch:2007} provide full details of the instrument, which is based around three separate Gas Proportional Counter (GPC) units filled with a P10 Argon-Methane mixture at 0.15 MPa (1.5 bar). In order to distinguish between characteristic X-ray lines of the major rock forming elements, and allow quantitative analysis of surface elemental abundances, one GPC carries a 4.5$\mu$m-thick Mg filter, one has a 6.3$\mu$m Al filter, and the third GPC is unfiltered.

The XRS was active during all three Mercury flybys, and a number of X-ray emission features were observed during the encounters.~\citet{Ho:2011} present a detailed analysis of several features of particular interest, attributing some to astrophysical sources, and some to the detection of suprathermal plasma electrons originating in Mercury's magnetosphere. ~\citet{Sch:2011} describe simulations and laboratory work on electron transport and acceleration in the Hermean magnetosphere and also interpret several X-ray features in the first and second flybys as the signature of suprathermal electrons. We give the times and durations of features considered by \citet{Ho:2011} (M1-E1, M2-E1, M2-E2 and M3-E1) in Table \ref{tab:events}, and define four additional features (M1-B1, M1-B2, M1-B3 and M2-B1) which we have also considered in the present work.

Our analysis concentrates on the sensitivity of the XRS to bright astrophysical X-ray sources, and began as work  conducted to support science planning activities for the Mercury Imaging X-ray Spectrometer (MIXS) instrument onboard ESA's {\it BepiColombo} mission to Mercury, with which the authors are affiliated~\citep{Fra:2010}. A detailed analysis of {\it MESSENGER} spacecraft attitude and XRS instrument pointing during the flybys has been conducted, generating a model of the passage of known bright X-ray sources through the field of view (FOV) as a function of time. Using measured instrument sensitivity data and a simple treatment of the angular variation of the collimator transmission function, the instrument count rates expected from these sources have been calculated and compared with measurements from the XRS. The final column of Table~\ref{tab:events} summarises our conclusion for the origin of each feature based on the present work.

\begin{table}[ht!]
\centering
\small
\begin{tabular}{l|l|l|l|l}
\hline
Event & Date & Start Time & Duration & Origin \\ 
 & & (UTC) & (sec) \\ \hline
M1-E1 & 14 Jan 08 & 18:59:49 & 180 & Electrons \& astrophysical\\
M1-B1 & 14 Jan 08 & 19:20:12 & 300 & Astrophysical\\
M1-B2 & 14 Jan 08 & 19:26:40 & 800 & Astrophysical\\
M1-B3 & 14 Jan 08 & 19:41:00 & 1040 & Astrophysical\\
M2-E1 & 6 Oct 08 & 08:35:08 & 60 & Electrons \& astrophysical\\
M2-E2 & 6 Oct 08 & 08:47:08 & 120 & Electrons \\
M2-B1 & 6 Oct 08 & 08:57:10 & 220 & Astrophysical \\
M3-E1 & 29 Sep 09 & 21:45:39 & 60 & Insufficient data, no clear \\
& & & & astrophysical source\\
\hline
\end{tabular} 
  \caption{XRS events considered in this work. Where features are discussed by~\citet{Ho:2011} we adopt the {\it M[x]-E[y]} labels defined by the authors; we assign additional features a label of the form {\it M[x]-B[y]}. In each case $x$ denotes the flyby number and $y$ identifies the feature in the XRS dataset for that encounter. The final column summarises our findings for the origin of each feature (see Section~\ref{sec:id} for details).}\label{tab:events}
\end{table}

In Section~\ref{sec:fov} we describe the reconstruction of instrument pointing and FOV location. Section~\ref{sec:rates} summarises the assumptions made in estimating the count rates generated by each object in the field, and in Section~\ref{sec:id} we consider the major XRS features from the first two flybys in detail, identifying correlations between observed X-ray peaks and the presence of bright sources in the field of view. In two features (M1-E1 and M2-E1) we find evidence to suggest that significant numbers of astrophysical source photons are contributing to the recorded flux, in addition to the proposed suprathermal electron element which we also support. The available XRS data for Flyby 3 are less detailed due to a safe-hold event which took place during the encounter, and a lower level of analysis has been conducted for that event, which we cover briefly in Section~\ref{sec:id3}.

\section{Spacecraft telemetry and field of view direction}\label{sec:fov}
The XRS observations considered in this work were obtained as calibrated data records from the Planetary Data System Geosciences Node hosted by Washington University in St. Louis (http://\-pds-\-geosciences.\-wustl.edu\-/\-miss\-ions\-/\-messenger/\-xrs.htm). Count rates measured by the three GPC channels are provided with 60 second time resolution in the periods around closest approach for each encounter. Full details of the calibrated data file structure are provided by~\citet{Har:2010}.

The attitude of {\it MESSENGER} was determined using the SPICE system produced by the NASA Navigation and Ancillary Information Facility (NAIF;~\citet{Act:1996}), which comprises spacecraft data and planetary ephemerides (both presented in data files referred to as ``kernels'') and the SPICE computational toolkit. For each flyby, the direction cosine matrix corresponding to the transformation between {\it MESSENGER} spacecraft body axes and J2000 coordinates was generated, for each attitude data point, with a temporal resolution of 0.36 seconds. The results were verified by combining these data with a spacecraft ephemeris in Mercury Solar Orbital coordinates, generated from the same kernels. In each case, the time, distance and location of closest approach agreed with published values~\citep{Ben:2010,Sla:2009b}. A list of the kernels used in this work can be found in \ref{sec:appendix}.

The {\it MESSENGER} XRS has a hexagonal FOV with a full-width-zero-maximum (measured from vertex to vertex) of 12$^{\circ}$, and a boresight aligned with the spacecraft +Z axis. The positions of the XRS FOV boresight and 6 perimeter points in the spacecraft frame are provided in the {\it msgr\_xrs\_v001.ti} kernel. These points were transformed into J2000 coordinates for each 0.36 second time interval, enabling precise mapping of the FOV onto the celestial sphere. In addition, spacecraft position and attitude data were imported into Analytical Graphics Inc.'s {\em Satellite Tool Kit} (STK)\footnote{http://www.stk.com}, and a sensor with the hexagonal FOV and orientation of the XRS instrument was reproduced to provide a time-resolved visualisation of the field position. The orientation of the FOV with respect to the planet in our model was found to be consistent with the state of the FOV\_STATUS flag in the XRS data. This flag characterises the contents of the FOV, and has five states: (0) no part of the planet in the field; (1) planet filling the field, at least part of footprint in sunlight; (2) planet filling the field, no portion in sunlight; (3) only part of the field includes the planet, at least one part of which is in sunlight, and (4) only part of the field includes the planet, with the imaged portion in darkness. The times at which the FOV made contact with the planet and crossed the dawn/dusk terminator in our simulation were in agreement with the status of this flag. 

\section{Source population \& count rate estimation}\label{sec:rates}
\subsection{The bright source sample}
The astrophysical X-ray sources included in this study were taken from the {\em ROSAT} All Sky Survey Bright Source Catalogue~\citep{Vog:1999} (hereafter RASSBSC). Using the electronic version of the catalogue hosted by the Leicester Database and Archive Service (LEDAS), the one hundred brightest sources as measured by the {\em ROSAT} Position Sensitive Proportional Counter (PSPC) were identified. 
The right ascension and declination of each object  ($\alpha_{2000}, \delta_{2000}$) were extracted from LEDAS and imported as a point source into STK. For a two hour period around the point of closest approach for each flyby, the identity of any of the sources within the FOV was determined, and the angular position of the source with respect to the instrument boresight recorded with 1 second time resolution. 

\subsection{Source count rates}
The {\it MESSENGER} XRS has a bandpass of 1 - 10 keV~\citep{Sch:2007}, significantly broader, and with a different effective area profile, than the 0.1 - 2.4 keV bandpass of the {\it ROSAT} PSPC~\citep{Pfe:1986}. Given the wide range of source spectra represented in the RASSBSC sample, it is not possible to adopt a constant scaling factor to convert from {\it ROSAT} to {\it MESSENGER} count rates. We therefore estimated the photon event rates expected in the XRS by modelling source spectra with the XSPEC X-ray spectral analysis package, using best-fit spectral parameters found in the literature. The sources, their type and count rates from RASSBSC and the count rate predicted for the {\it MESSENGER}  XRS (along with the references from which we obtained spectral parameters) are indicated in Table~\ref{tab:sources}, while the spectral hardness of the sources is indicated by the Al:unfiltered and Mg:unfiltered channel count rate ratios shown in Figure~\ref{fig:hard_ratio}.

Ground calibration data providing the effective areas of the Al, Mg and unfiltered XRS channels were available from the PDS\footnote{file XRS\_GPC\_MODELLED\_RESPONSE} and were used in this work. The variation in the collimator transmission function with angle was approximated as a triangular profile in which the observation efficiency was set to 1.0 at the centre of the field (representing on-axis sources), falling linearly to zero at an angle of 6$^{\circ}$ with respect to the instrument boresight.

\begin{figure}[ht!]
\centering
 \includegraphics[width=4in]{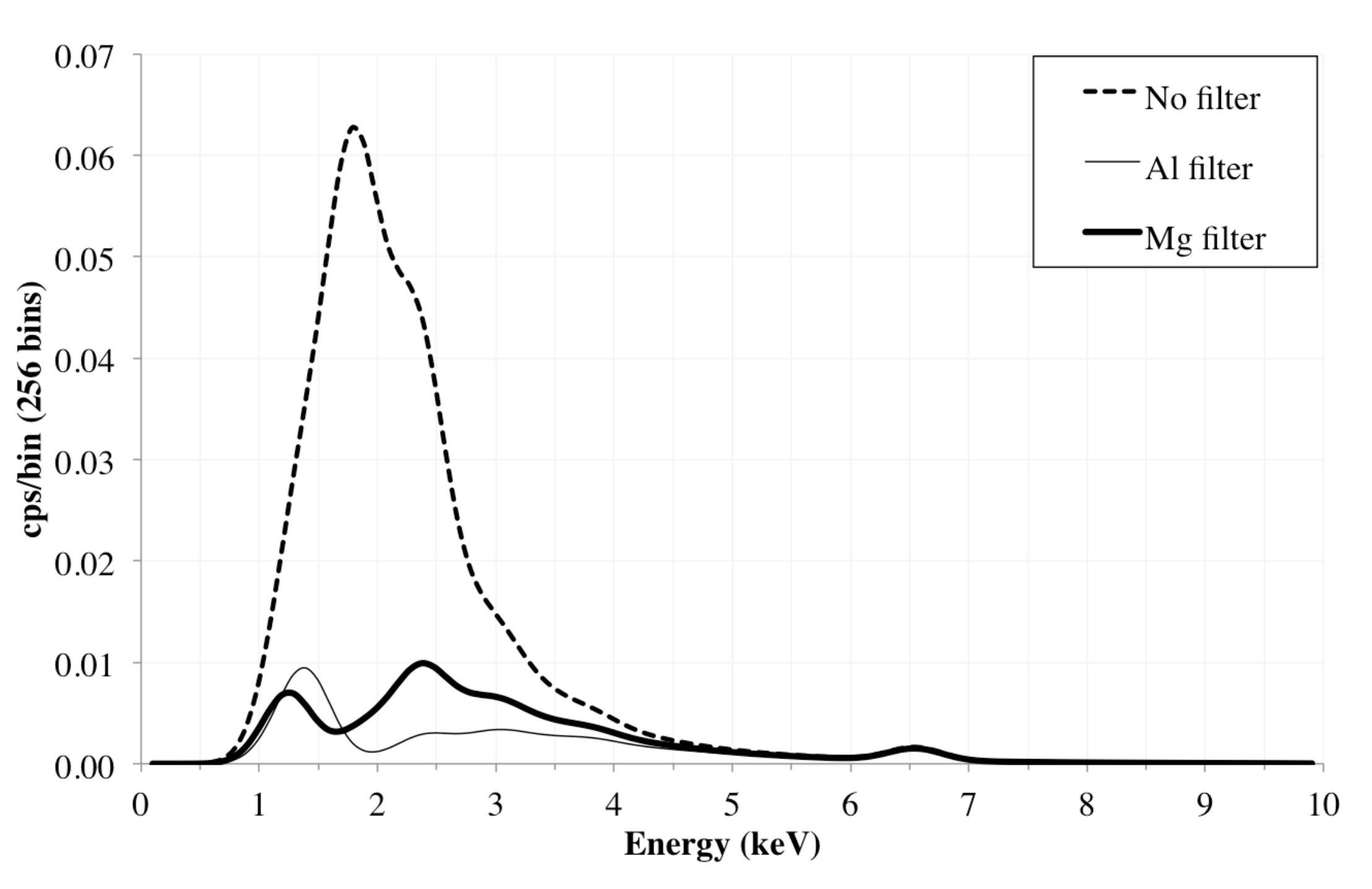}
 \caption{Spectrum of the Cas-A supernova remnant produced by our model, using the broken power law + emission line model of \citet{All:1997} and adopting the same 256 channel binning of \citet{Sch:2007}.}
\label{fig:casa}
\end{figure}

\begin{figure}[t!]
  \centering
  \includegraphics[width=4.5in]{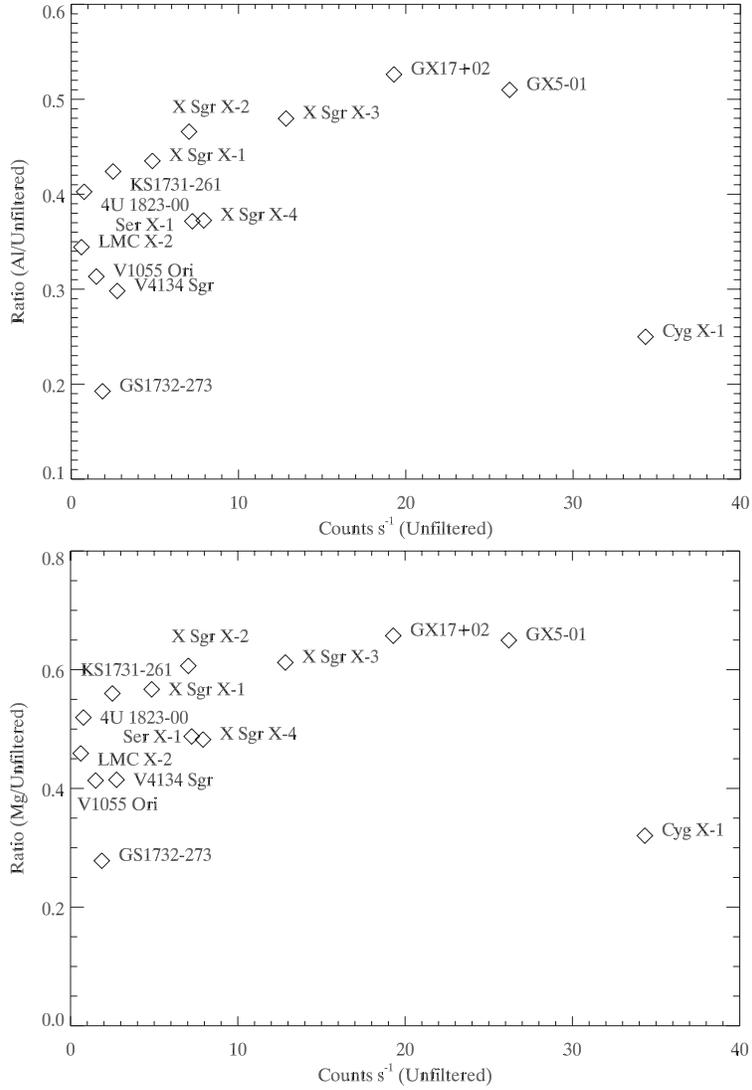}
  \caption{Hardness ratios for the sources listed in Table~\ref{tab:sources}. Top panel: on-axis unfiltered channel count rate versus the ratio of the on-axis Al and unfiltered channel rates. Bottom panel: on-axis unfiltered channel count rate versus the ratio of the on-axis Mg and unfiltered channel rates. Note the particularly soft ratio of Cyg X-1, consistent with the black hole nature of this object as discussed by~\citet{Whi:1984}.}
  \label{fig:hard_ratio}
\end{figure}

Observations of Cas-A have been used to calibrate the instrument during flight, and \citet{Sch:2007} show a spectrum of the object (Figure 13 in that work). We have modelled the Cas-A source spectrum using the broken power law + emission line model of \citet{All:1997}, and we find very close agreement between the predicted XRS observation (Figure~\ref{fig:casa}) and the data of \citeauthor{Sch:2007}.

\section{Astrophysical sources in XRS data}\label{sec:id}
Table~\ref{tab:sources} lists the sources found to enter the XRS FOV in the time periods of interest. Fifteen of the sources appear during the two hour time windows around each of the first two flybys which we consider in detail in this paper; one further source (Cyg X-1) is relevant to the discussion of Flyby 3 in Section~\ref{sec:id3}. 

\begin{table}[ht!]
\centering
\small
\begin{tabular}{r|l|l|l|r|r|r|r}
\hline
BSC & 1RXS-J Number & Name & Type & PSPC & \multicolumn{3}{c}{XRS} \\ 
 & & & &  & Un & Al & Mg\\ \hline
5$^{[1]}$	&  182340.5-302137 	&	X Sgr X-4	& LMXB	&  127.0 	& 7.92 &    2.95 &    3.82	\\
8$^{[1]}$	& 183956.9+050203	&	Ser X-1	& LMXB	& 69.62	& 7.24 &    2.69 &    3.53	\\
11$^{[6]}$	& 195821.9+351156	&	Cyg X-1	& HMXB & 61.25 & 34.33 &    8.58 &   11.00\\
13$^{[1]}$	& 061707.4+090812 	& 	V1055 Ori	& LMXB	& 48.66 	& 1.50 &    0.47 &    0.62\\
16$^{[1]}$ 	& 175840.1-334828	&	V4134 Sgr	& LMXB	& 31.08	& 2.75 &    0.82 &    1.14 \\
17$^{[1]}$	& 180132.3-203132	&	X Sgr X-3	& LMXB	& 30.92	& 12.84 &    6.16 &    7.86 	\\
18$^{[2]}$	& 173602.0-272541	&	GS1732-273	& LMXB	& 29.79	& 1.87 &    0.36 &    0.52	\\
21$^{[1]}$	& 181601.2-140213	&	GX17+02	& LMXB & 24.82	& 19.29 & 10.15 & 12.68	 	\\
27$^{[1]}$	& 174755.8-263352 	& 	X Sgr X-1	& LMXB & 19.2 	& 4.85 &    2.11 &    2.75	 \\
32$^{[1]}$	& 180108.7-250444	&	GX5-01	& LMXB & 17.78	& 26.20 &   13.36 &   17.02 	\\
35$^{[3]}$	& 173413.0-260527 	& 	KS1731-261	& LMXB & 14.09 	& 2.50 &    1.06 &    1.40\\
36$^{[4]}$	& 064509.3-164241 	& 	Sirius B	& WD & 12.73 	& N/A  & N/A & N/A \\
37$^{[1]}$	& 052027.6-715755	&	LMC X-2 & LMXB 	& 12.17	& 0.61 &    0.21 &    0.28 \\
75$^{[5]}$	& 162909.4+780439 	& 	WD1631+78 & WD & 5.66 	& N/A & N/A & N/A \\
76$^{[1]}$	& 182522.1-000038 	& 	4U 1823-00	& LMXB & 5.66	& 	0.77 &    0.31 &    0.40 \\
87$^{[1]}$	& 181431.6-170917	&	X Sgr X-2	& LMXB & 4.88	& 7.04 &    3.28 &    4.27	\\
\hline
\end{tabular} 
  \caption{Sources in the ROSAT Bright Source Catalogue which appear in the {\it MESSENGER} XRS FOV during the Mercury flybys. Column 1 indicates the brightness ranking in the RASSBSC (i.e. X Sgr X-4 is the fifth brightest source in the catalogue). PSPC and XRS columns give the measured PSPC count rate and estimated XRS count rate in the three channels for sources on-axis. Numbers in square brackets in column 1 indicate the references used to construct the model source spectra: [1]~\citet{Chr:1997}; [2]~\citet{Yam:2004}; [3]~\citet{Bar:1998}; [4]~\citet{Hol:1998}; [5]~\citet{Odw:2003}.}\label{tab:sources}
\end{table} 

The majority of the sources are low mass X-ray binary (LMXB) systems.~\citet{Whi:1985} describe a function of the form
\begin{equation}
\label{eq:usc}
N_{source}(E)=KE^{-\Gamma}\exp(-E/k_BT_{th}),
\end{equation}
which represents the unsaturated Comptonised spectrum of cool photons upscattered on hot photons, and which has been found to provide a good approximation to LMXB spectra. A survey of bright LMXBs observed with the {\em Einstein} observatory was conducted by~\citet{Chr:1997} who present spectral parameters of this form for 49 LMXBs, including 11 of the objects of interest in the current work. We have used these parameters to model the LMXB spectra as observed by {\it MESSENGER}.  

Two of the objects in our list (Sirius B and WD1631+78) are white dwarf stars. As a class of object, white dwarfs can exhibit a range of complex spectra due to heavy elements suspended in the atmosphere through the effects of radiation pressure. However, for the purposes of this study, it is sufficient to approximate the white dwarf spectrum as a blackbody with effective temperature $T_{eff}$.~\citet{Hol:1998b} give  $T_{eff}=24,790$ K for Sirius B, and $T_{eff} = 44,560$ K for WD1631+78. Both objects are therefore capable of generating X-ray photons above the 0.1 keV lower limit of the RASSBSC bandpass, but are not significant sources of X-rays with energies above the 1.0 keV lower limit of the {\it MESSENGER} XRS. Spectra for Sirius B presented in e.g.~\citet{Pea:2000} and ~\citet{Beu:2006} support this conclusion.~\citet{Odw:2003} list WD1631+78 as a source of ``hard'' ($>0.5$ keV) X-rays, probably from its unresolved dM4e companion star. Again, the indicated flux levels are extremely low in the {\it MESSENGER} XRS bandpass, and hence we ignore these objects in the model.

\begin{figure}[t!]
  \centering
  \includegraphics[width=2.9in]{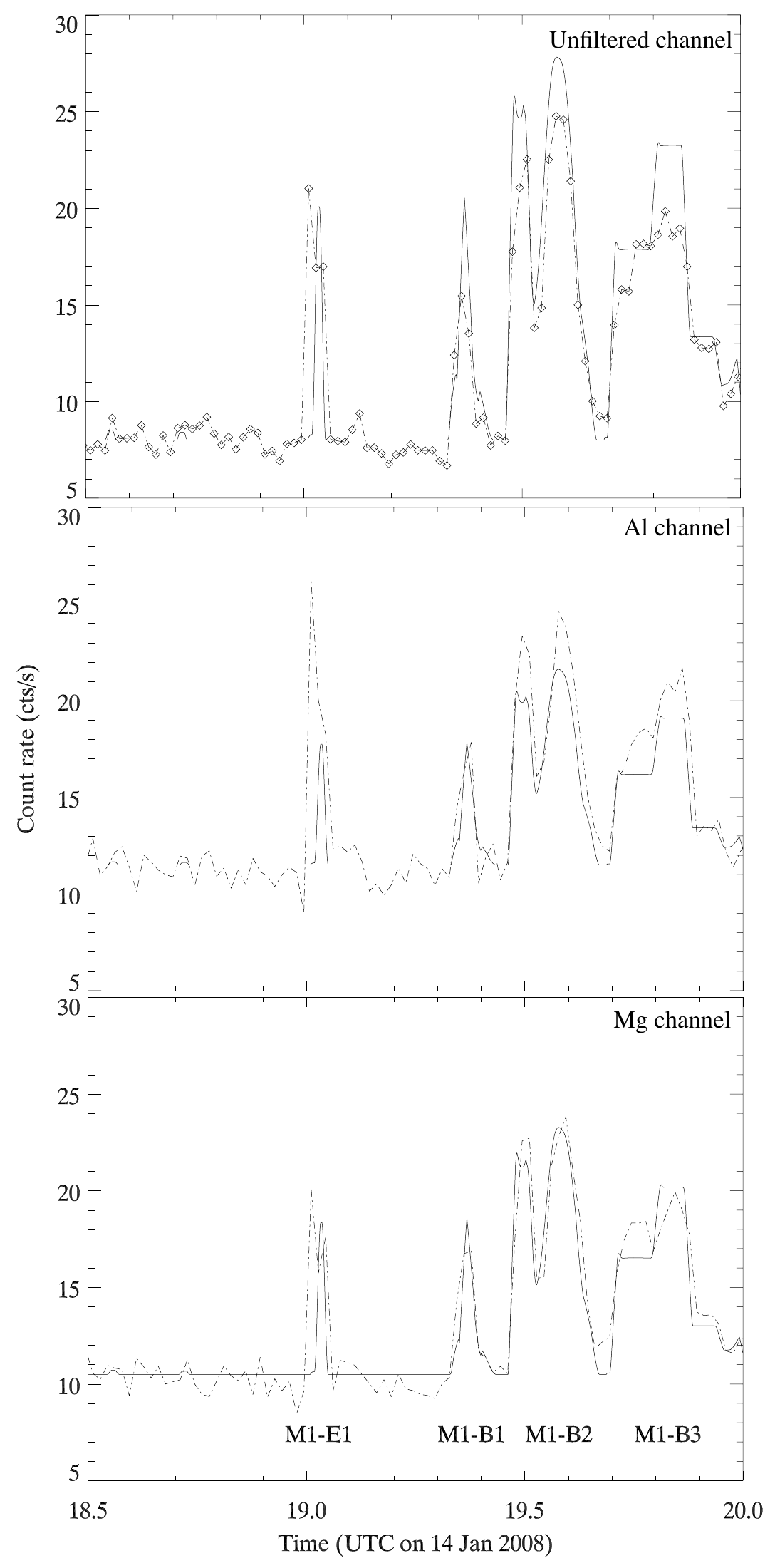}
  \caption{XRS observations (dotted lines) compared to the predicted count rates from astrophysical sources (solid lines) for flyby 1. XRS data are shifted by +50 seconds to match the timing of M1-B2, and the model smoothed to match the 60 second time resolution of the observations before an offset of 8.2, 11.5 and 10.6 cts s$^{-1}$ is added to approximate the XRS background level in the unfiltered, Al and Mg channels respectively. The labels identify the features considered in this work. Individual XRS data bins are indicated by diamond symbols in the unfiltered channel plot.}
  \label{fig:flyby1}
\end{figure}

\begin{figure}[ht!]
  \centering
  \includegraphics[width=2.9in]{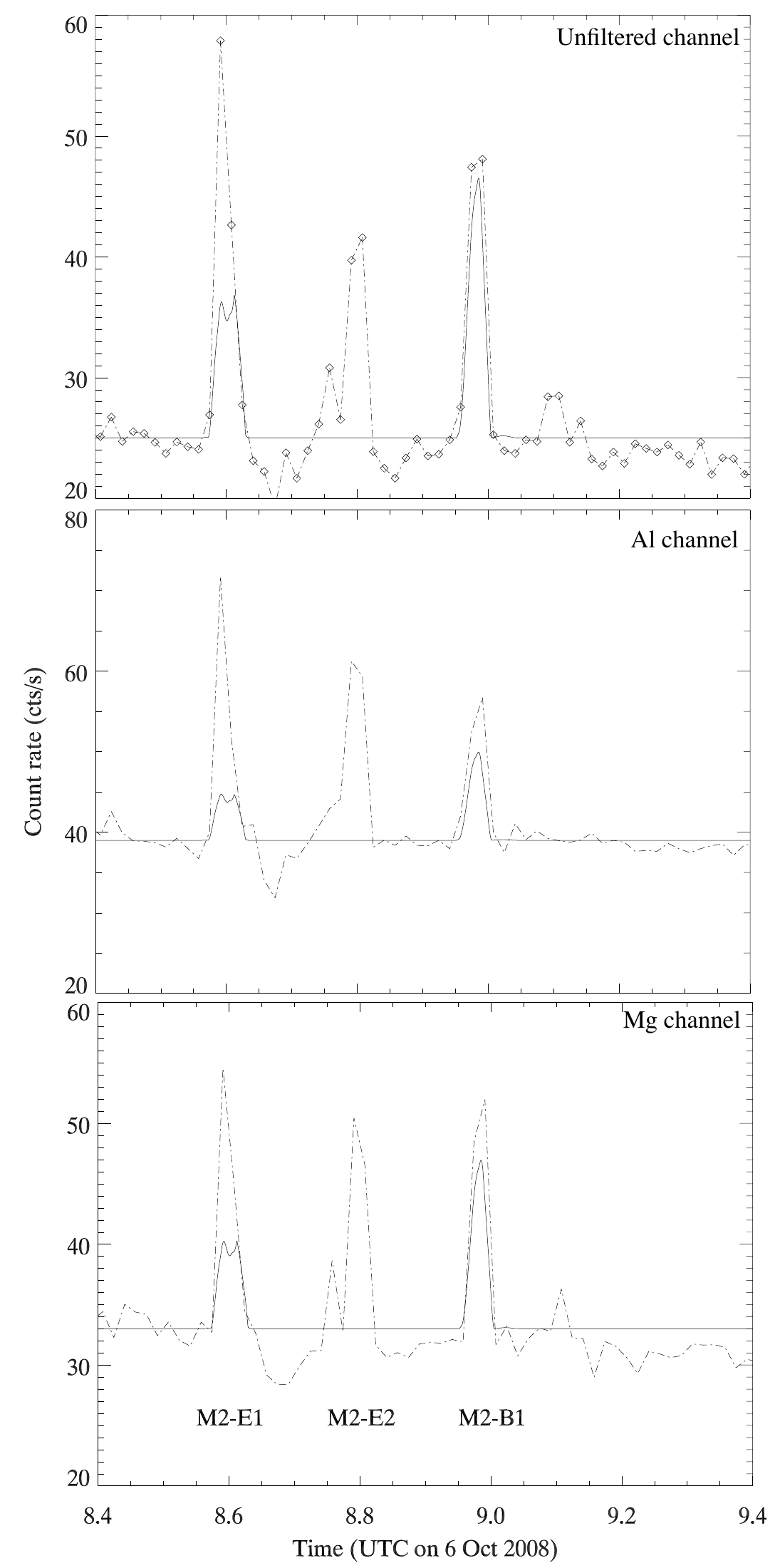}
  \caption{XRS observations (dotted lines) compared to the predicted count rates from astrophysical sources (solid lines) for flyby 2. XRS data are shifted by +20 seconds to match the timing of M2-B1, and the model smoothed to match the 60 second time resolution of the observations before an offset of 25.9, 40.6 and 34.1 cts s$^{-1}$ is added to approximate the XRS background level in the unfiltered, Al and Mg channels respectively. The labels identify the features considered in this work. Individual XRS data bins are indicated by diamond symbols in the unfiltered channel plot.}
  \label{fig:flyby2}
\end{figure}

Table~\ref{tab:sources} shows the count rates expected in the XRS from each source. These values, combined with instrument response data, can be used to predict the XRS event rates due to astrophysical sources. Figures~\ref{fig:flyby1} and~\ref{fig:flyby2} show the count rate recorded in the three XRS channels (dotted lines) as a function of time as the FOV moves across the sky, compared with the count rates predicted by our model (solid line) for the first two flybys. The model has been smoothed to match the 60 second time resolution of the observational data. An offset of +50 seconds has been added to the XRS data for flyby 1, and +20 seconds for flyby 2. This correction is necessary to provide the optimum temporal correlation between observation and the model and is justified as it is smaller than the temporal resolution of the observed data. The magnitude of the correction was determined using the times of observed features that are of purely astrophysical origin. Instrument background levels were estimated from the average XRS count rate in each channel, evaluated over a $\sim$ three hour period immediately prior to each flyby (15:47:49 UTC to 18:45:49 UTC on 14th January 2008 for flyby 1, 05:25:08 UTC to 08:24:08 UTC on 6th October 2008 for flyby 2). The resulting background values were 8.2, 11.5 and 10.6 cts s$^{-1}$ for the unfiltered, Al and Mg channels during flyby 1, and 25.9, 40.6 and 34.1 cts s$^{-1}$, for the unfiltered, Al and Mg channels during flyby 2, respectively. 

\begin{sidewaystable}[H!]
\centering
\footnotesize
\begin{tabular}{l|l|lllr|rrr|rrr}
\hline
{\bf Event} & Star & $t_{entry}$ & $t_{exit}$ & $t_{max}$ & $\theta_{min}$ & \multicolumn{3}{|c|}{Peak count rate} & \multicolumn{3}{|c}{Total  (\%)}\\ 
 & & (UTC) & (UTC) & (UTC) & ($^{\circ}$)& $C_{un}$ & $C_{al}$ & $C_{mg} $ & $T_{un}$ & $T_{al}$ & $T_{mg} $\\ \hline
M1-E1 	& GX17+02 & 19:01:38 & 19:02:28 &  19:02:00 & 0.14 & 18.83 & 9.91 & 12.38 & 35.25 & 18.17 & 33.75\\
		& X Sgr X-2 & 19:01:48 & 19:02:23 & 19:02:06 & 2.75 & 3.83 & 1.78 & 2.32 & 5.44 & 2.49 & 4.81\\ 
		& Ser X-1 & 19:01:02 & 19:01:09 & 19:01:05 & 4.94 & 1.30 & 0.48 & 0.63 & 0.52 & 0.19 & 0.37\\
		& 4U1823-00 & 19:01:06 & 19:01:28 & 19:01:17 & 1.62 &0.56 & 0.23 & 0.29 & 0.46 & 0.18 & 0.35\\		\hline
		
M1-B1	& GX5-01 & 19:21:33 & 19:23:26 & 19:21:33 & 2.93 & 13.45 & 6.86 & 8.74 & 105.50 & 66.77 & 62.63\\
		& X Sgr X-1 & 19:20:25 & 19:22:23 & 19:20:30 & 1.75 & 3.44 & 1.50 & 1.95 &  23.16 & 12.50 & 12.00\\
		& X Sgr X-4 & 19:23:42 & 19:25:29 & 19:23:42 & 3.60 & 3.18 & 1.19 & 1.54 & 20.46 & 9.46 & 9.02\\
		& GS1732-273 & 19:20:13 & 19:21:23 & 19:20:30 & 3.45 & 0.80 & 0.15 & 0.22 &  3.24 & 0.77 & 0.82\\
		& KS1731-261 & 19:20:17 & 19:20:51 & 19:20:30 & 4.32 & 0.71 & 0.30 & 0.40 &  1.89 & 1.00 & 0.97 \\ \hline
		
M1-B2	& GX17+02 & 19:30:45 & 19:37:30 & 19:34:11 & 1.07 & 15.86 & 8.35 & 10.43 &  55.33 & 36.40 & 43.62 \\
		& GX5-01 & 19:28:13 & 19:30:41 & 19:30:11 & 2.78 & 14.10 & 7.19 & 9.62 &  26.84 & 17.11 & 20.91\\
		& X Sgr X-2 & 19:30:34 & 19:39:45 & 19:35:56 & 0.38 & 6.60 & 3.08 & 4.00 &  22.92 & 13.35 & 16.67\\ 
		& X Sgr X-3 & 19:28:21 & 19:39:38 & 19:30:41 & 1.25 & 10.19 & 4.89 & 6.24 &  18.31 & 10.99 & 13.44 \\ \hline
		
M1-B3	& GX5-01 & 19:47:51 & 19:52:36 & 19:48:25 & 2.77 &14.16 & 7.22 & 9.20 &  47.80 & 34.34 & 40.61\\ 
		& X Sgr X-3 & 19:42:12 & 19:48:33 & 19:42:36 & 2.83 & 6.80 & 3.26 & 4.16 &  30.21 & 20.42 & 24.18\\
		& X Sgr X-4 & 19:41:44 & 19:58:20 & 19:52:35 & 2.53 & 4.59 & 1.71& 2.22 &  20.57 & 10.80 & 12.98\\
		& X Sgr X-2 & 19:42:11 & 19:48:11 & 19:42:36 & 2.17 & 4.50 & 2.10 & 2.73 &  17.21 & 11.30 & 13.65\\
		& V4134 Sgr & 19:52:28 & 19:58:20 & 19:57:01 & 1.35 & 2.14 & 0.64 & 0.89 &  6.12 & 2.57 & 3.32\\
		& X Sgr X-1 & 19:52:14 & 19:52:22 & 19:52:20 & 5.59 & 0.35 & 0.15 & 0.20 &  0.04 & 0.02 & 0.03\\ \hline
		
M2-E1	& X Sgr X-3 & 08:35:43 & 08:37:14 & 08:37:14 & 0.97 & 10.77 & 5.17 & 6.59 &  22.24 & 13.20 & 23.07\\
		& GX17+02 & 08:34:52 & 08:36:06 & 08:35:17 & 2.32 & 11.85 & 6.24 & 7.79 &  18.00 & 11.72 & 20.06 \\
		& X Sgr X-2 & 08:35:03 & 08:37:29 & 08:36:01 & 3.83 & 2.57 & 1.20 & 1.56 &  9.08 & 5.23 & 9.34\\ 
		& GX5-01 & 08:37:02 & 08:37:16 & 08:37:16 & 4.87 & 5.03 & 2.57 & 3.27 &  1.99 & 1.26 & 2.19\\
		& 4U 1823-00 & 08:34:16 & 08:34:34 & 08:34:25 & 4.18 & 0.24 &  0.09 & 0.12 & 0.12 & 0.06 & 0.10\\ \hline

M2-B1	& GX5-01 & 08:57:53 & 08:59:47 & 08:59:31 & 0.44 & 24.3 & 12.39 & 15.79 & 74.21 & 58.73 & 77.95\\
		& X Sgr X-4 & 08:57:35 & 08:58:28 & 08:57:53 & 3.97 & 2.70 & 1.01 & 1.30 & 3.95 & 2.29 & 3.08\\
		& KS1731-261 & 08:59:11 & 08:59:24 & 08:59:20 & 5.35 & 0.28 & 0.12 & 0.16 &  0.12 & 0.08 & 0.11\\
		& 4U1823-00 & 09:00:38 & 09:01:00 & 09:01:00 & 4.54 & 0.19 & 0.08 & 0.10 & 0.11 & 0.07 & 0.09\\
\hline
\end{tabular} 
  \caption{Astrophysical source contributions to each event in Flybys 1 and 2 (except M2-E2 which contains no astrophysical component). For each source, the times of entry into / exit from the field of view are provided ($t_{entry, exit}$), as is the time at which the source contribution is at a maximum ($t_{max}$), corresponding to the point in time at which the source was closest to the centre of the field of view (off-axis angle $\theta_{min}$). For each GPC channel, the peak count rate $C$ and the fraction of the total observed event $T$ represented by the source, after background subtraction, is given.}\label{tab:event_contrib}
\end{sidewaystable}
\clearpage

For any given feature, the model rates in the Al and Mg channels depicted in Figures~\ref{fig:flyby1} and~\ref{fig:flyby2} are significantly lower than those in the unfiltered channel at the same instant, as expected on the basis of the filter transmission properties. However, for those features which are unambiguously attributed to astrophysical sources, we find that the difference between the observed rates in the unfiltered channel and those in the Al and Mg channels, after background subtraction, is smaller than predicted by the model. For example, the second (higher) peak in M1-B2 has background-subtracted observed count rates of approximately 16.8, 13.5 and 13.4 cts s$^{-1}$ in the unfiltered, Al and Mg channels respectively, while the model count rates are approximately 19.8, 10.5 and 12.4 cts s$^{-1}$, so that in the model, the unfiltered peak has almost twice the intensity of the filtered peaks. We suggest that this larger-than-observed difference between unfiltered and filtered channel rates in our model is due to the fact that the model excludes the generation of X-ray fluorescence photons in the filters. We find that the number of fluorescence photons predicted for a signal of the strength of M1-B2 is $\sim 6-8$ which, when added to the directly detected events in the model, is of the correct order to reduce the difference between filtered and unfiltered rates, making their ratios more similar to those of the three channels in the observed signal. Note, however, that fluorescence photons cannot explain the much more significant under-prediction of flux in the modelled M1-E1 and M2-E1 features, or the fact that the observed Al count rates {\em exceed} the unfiltered values in those features. A more detailed discussion of the production of fluorescence photons in the filters is provided in Section~\ref{sec:m2e1disc}, in the context of the under-prediction of count rates for M2-E1.
 
The results of our model reveal significant correlations with observed features. We identify the contributing sources in Table~\ref{tab:event_contrib}, and discuss the events and sources below. Note that while the quoted value of $C_{max}$ in Table~\ref{tab:event_contrib} corresponds to the count rate evaluated for the one-second time bin centred on $t_{max}$,  the plotted model data in Figures~\ref{fig:flyby1} and~\ref{fig:flyby2} have been smoothed to match the 60 second time resolution of the observations, which reduces the peak height and broadens the feature in the plots. We find that the model levels in the unfiltered channel for M1-E1 appear to be ~20\% higher than observed (as reflected in Figure~\ref{fig:flyby1} and Table~\ref{tab:event_contrib}) possibly due to over-estimation of, or change in, the background level at this time, but the agreement in Al and Mg channels is considerably better, and the similarity between XRS observations and the model is observed in all three channels during Flybys 1 and 2. 

{\it Unless otherwise stated, count rates quoted for objects in the following sections refer to those measured in the unfiltered channel.}

\subsection{M1-E1}\label{sec:m1e1}
This feature comprises at least two peaks, producing the stepped profile visible in Figure~\ref{fig:flyby1}. Our model shows a significant astrophysical component for $\sim64$ seconds starting near 19:00:50 UTC. The principal source is the LMXB GX17+02 contributing a maximum of 18.8 cts s$^{-1}$, with additional flux from Sgr X-2 and minor contributions  from Ser X-1 and 4U 1823-00. We discuss this event in detail in Section~\ref{sec:m1e1disc}

\subsection{M1-B1}
The dominant source is GX5-01 ($\lesssim 4.8$ cts s$^{-1}$) followed by Sgr X-1 and Sgr X-4, each generating $\lesssim 1.2$ cts s$^{-1}$. GS1732-273 also appeared in the field at this time, generating $\lesssim 0.3$ cts s$^{-1}$. The sharp, stepped shape of this feature contrasts to other observations and is due to the fact that the sources were observed emerging from occultation behind the limb of Mercury following closest approach, so that the source is first observed near the centre of the FOV rather than entering from the field edge as would be the case for an unobstructed slew manoeuver.

\subsection{M1-B2}\label{sec:m1b2}
At this time in the encounter, the XRS FOV was passing over a number of bright sources. The principal contributions in the first peak of this feature come from Sgr X-3 (2.1 cts s$^{-1}$), GX17+02 (3.0 cts s$^{-1}$), GX5-01 (5.3 cts s$^{-1}$) and Sgr X-2 (1.4 cts s$^{-1}$). The second peak is a revisit of these sources, except for GX5-01 which is not observed again.

\subsection{M1-B3}
The dominant sources in this feature are GX5-01 (peak count rate 5.0 cts s$^{-1}$), Sgr X-3 (2.4 cts s$^{-1}$), Sgr X-4 (1.7 cts s$^{-1}$) and Sgr X-2 (1.6 cts s$^{-1}$). Minor contributions come from V4134 Sgr (0.8 cts s$^{-1}$) and Sgr X-1 (0.1 cts s$^{-1}$). The XRS field scanned this area over an extended period of time, explaining the broad shape of this feature.

\subsection{M2-E1}
We find that this feature coincides very closely with the presence of two bright sources and two fainter objects in the FOV. The feature begins at 08:34:52 with the entry of GX17+02 into the field, producing a maximum of 4.2 cts s$^{-1}$ at 08:35:17 UTC. 4U1823-00 and Sgr X-2 are minor contributors to the peak, at $<0.08$ and $<0.9$ cts s$^{-1}$ respectively. At 08:35:43, as GX17+02 left the field, Sgr X-3 entered; the changing sum of fluxes from these two bright objects produces the local minimum in the centre of the modelled feature. GX5-01 entered the field at 08:37:02, generating a maximum of 1.8 cts s$^{-1}$. 

At this time, the slew rate of the field in the J2000 frame was very low. Sgr X-3 entered the field at an angular rate of less than 0.1$^{\circ}$ s$^{-1}$. 90 seconds later, the object was within 1$^{\circ}$ of the FOV centre generating $\sim 3.8$ cts s$^{-1}$, and the slew rate was less than 0.01$^{\circ}$ s$^{-1}$. This feature occured immediately before the XRS FOV passed onto the planet, and the astrophysical sources were lost as they were occulted by the planet's limb. Sgr X-3 was occulted at 08:37:14, and GX5-01 was lost two seconds later. We discuss this event in detail in Section~\ref{sec:m1e1disc}

\subsection{M2-E2}
Our attitude data show that at the time of the M2-E2 event, the XRS FOV was entirely filled with a region of the planet near the terminator, in agreement with the FOV\_STATUS flag in the telemetry files. Therefore there can be no astrophysical contributions to the events contained within this feature. 

\subsection{M2-B1}
This feature is dominated by the presence of GX5-01 which contributes a maximum of 8.6 cts s$^{-1}$,  and Sgr X-3 (4.3 cts s$^{-1}$). Weaker contributions come from Sgr X-1 and Sgr X-4, each at approximately 1.0 ct s$^{-1}$ when closest to the FOV centre. Very low count rates ($<0.1$ cts s$^{-1}$) are contributed by 4U1823-00 and KS1731-261.

\section{Flyby 3 data}\label{sec:id3}
Our study has concentrated on the first and second {\it MESSENGER}  flybys of Mercury. During the third flyby on 29th September 2009, the spacecraft experienced a safe-hold event immediately before the time of closest approach. Observations of significant scientific value were obtained during the inbound phase of the encounter (e.g.~\citet{Sla:2010,Bak:2011}), but the XRS data are relatively featureless until the appearance of an event referred to as M3-E1 by~\citet{Ho:2011}, beginning at 21:45:39 UTC and lasting for 60 seconds. Inspection of calibrated and experiment XRS data records for this flyby shows that the final data point in those files covers precisely the same period of time.

Our data show that at 21:48:33 UTC, Cyg X-1 entered the field of view and remained there for 55 seconds, but this time shift is too large to be permitted on the basis of the temporal resolution of the XRS data. The angular separation between XRS boresight and Cyg X-1 was of order $\sim90^{\circ}$ at the time of the peak, and so no flux from the object can have leaked through the collimator. Thus, no astrophysical source can be identified to match the appearance of M3-E1; the nature of the observational data make more detailed analysis difficult to achieve.

\section{Discussion}
The majority of the XRS features considered in this work are clearly generated by astrophysical sources in the FOV; they are not considered by~\citet{Ho:2011} for this reason, and we have included them here for completeness only. Of greater interest are the results for M1-E1 and M2-E1. Both are discussed by~\citet{Ho:2011} who propose suprathermal electrons from the Hermean magnetosphere as their origin. We find that there is a significant correlation between these events and the count rate behaviour expected on the basis of our source modelling. 

The spectra shown in this section were extracted from the PDS-sourced files using custom-written IDL code. The data, supplied as counts per bin per integration period, were converted to counts per 60 seconds as a function of energy using telemetry along with calibration data supplied at the PDS Geosciences Node. The plots were smoothed using a boxcar filter with a smoothing window of 23 channels, equivalent to the quotient of the energy resolution of XRS at 5.9 keV (880 eV;~\citet{Sch:2007}) and the energy width of the spectral bins (37.9 eV). We note that this filter is too wide at low energy, but is adequate for the current analysis.

\subsection{M1-E1}\label{sec:m1e1disc}
This feature has an unusual shape consisting of two closely-spaced components, leading to a peak followed by a plateau of slightly lower count rate. The feature occurred while the spacecraft was executing a large slew near the time of closest approach. The FOV made contact with the planet's limb at approximately 19:02:35 UTC, and was completely filled with the planet by 19:03:20 UTC. The time of closest approach was 19:04:13 UTC.
\begin{figure}[t!]
  \centering
  \includegraphics[width=4.5in]{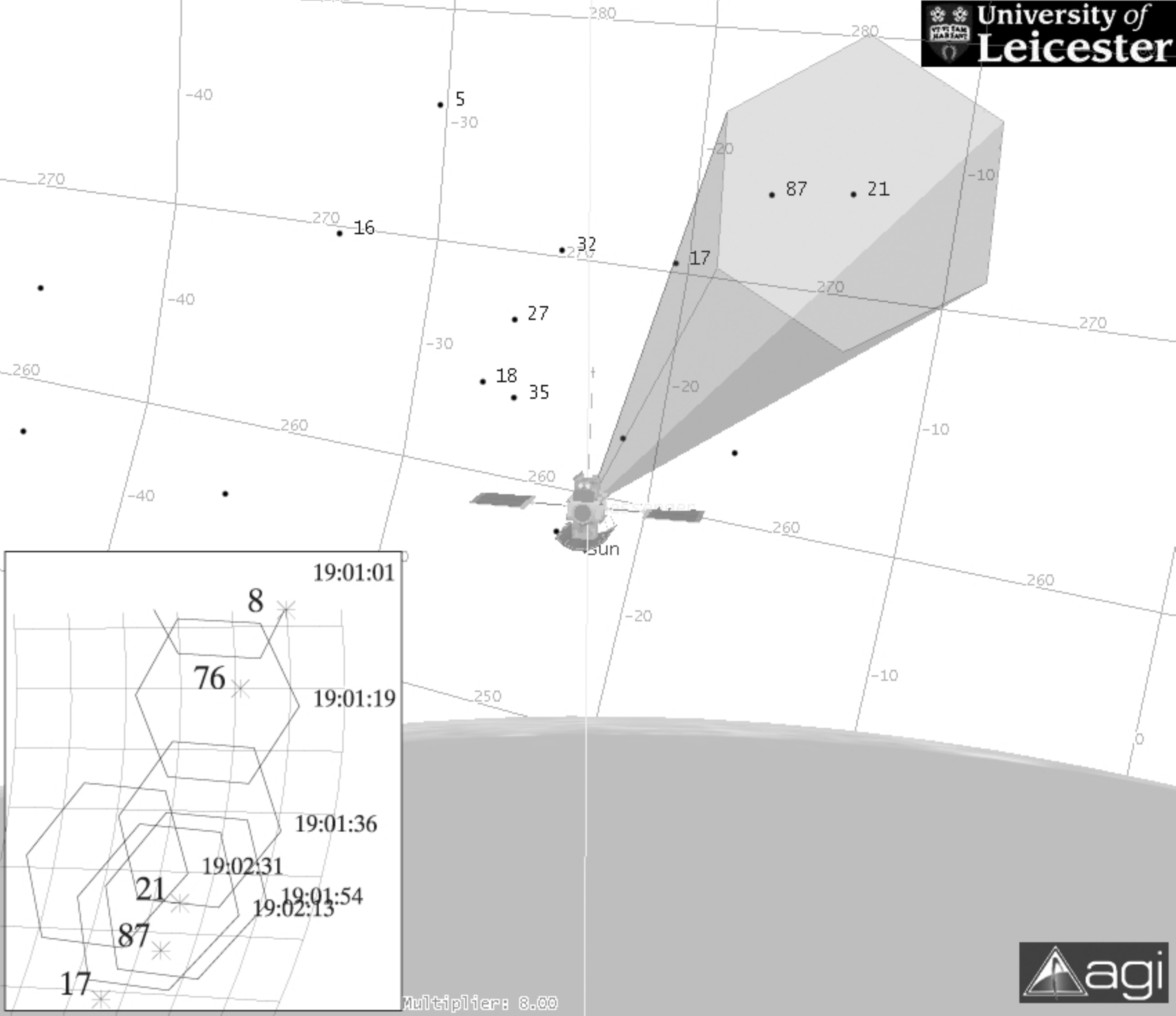}
  \caption{Orientation of the XRS FOV at 19:02:00 UTC on 14th Jan 2008, during the M1-E1 event. This time corresponds to the maximum astrophysical source count rate indicated in our model. Inset: the location of the FOV over time during the M1-E1 event, with times indicated in UTC. Asterisk symbols indicate the position of bright sources, the numbers corresponding to those given in Table~\ref{tab:sources}.}
  \label{fig:m1e1stk}
\end{figure}

~\citet{Ho:2011} define M1-E1 as an event of 180 s duration beginning at 18:59:49 UTC. GX17+02 is responsible for the major peak predicted by our model; it entered the field at 19:01:38 UTC and exited at 19:02:28 UTC, entirely within the period covered by M1-E1. Our model indicates a peak count rate of 6.6 cts s$^{-1}$ at 19:02:00 UTC, so that a significant number of events recorded by the XRS within the 180 sec period of M1-E1 are from this source. We note that~\citet{Ho:2011} refer to an astrophysical source event of undetermined origin following the M1 closest approach, but since GX17+02 appears within the 180 sec timeframe of M1-E1, we assume that this is not the event they refer to (conversely, M1-B1, B2 and B3 {\em are} astrophysical sources observed following closest approach).

Our model fails to reproduce the first component of M1-E1 (the initial rise from background, through to the primary peak). Any astrophysical source proposed to explain this feature must be of substantial X-ray brightness, and no other sufficiently bright objects exist in the RASSBSC at this location; a search of the SIMBAD database\footnote{http://simbad.u-strasbg.fr/simbad/} also provides no additional candidates. We have examined the next 200 brightest objects in the RASSBSC for sources at appropriate positions to explain this feature, but all other sources in the area are too faint to explain the intensity of the feature, and no combination of count rates from these sources allows us to replicate the shape of M1-E1. 

A search of the WGACAT catalogue \citep{Whi:1994} was conducted to locate any notable sources present in the first $\sim$1 minute of the event. WGACAT is a point source catalogue generated from all ROSAT PSPC pointed observations, and hence count rates can differ significantly between the RASSBSC and WGACAT entries for a specific object if, for example, pointed observations were made of a target during an episode of increased activity. The cataclysmic variable AM Her (1RXS J181613.8+495203) was targeted in a pointed observation between 12th - 13th April 1991 when it was found to be in a high state; the recorded count rate of 41.6 cts s$^{-1}$ (compared to 0.13 cts s$^{-1}$ in the RASSBC) makes it the eighth brightest WGACAT source. We find that AM Her entered the field of view of the MESSENGER XRS on January 14th 2008 at 18:59:16 UTC, 33 seconds before the onset of M1-E1. It was closest to the field centre ($3.03^{\circ}$) at 18:59:27 UTC, and remained in the field for $\sim 21$ sec,  exiting at 18:59:38 UTC. 

Based on optical monitoring data from the American Association of Variable Star Observers (AAVSO), AM Her was in a high state during the first Mercury flyby (the data suggest that by the time of the second flyby, the object was in a low state). Although we have been unable to identify X-ray observations of the object made during the time around the first flyby,~\citet{Sch:2006} discuss {\em XMM Newton} observations of AM Her in July 2005 when the object was in an intermediate (close to high) state, as reflected by the AAVSO data for this period.  The spectrum presented by~\citeauthor{Sch:2006} (Figure 2 in that work) shows that a flux of $\sim 5$ counts s$^{-1}$ keV$^{-1}$ at 1 keV, falling to $\sim 0.1$ counts s$^{-1}$ keV$^{-1}$ at 10 keV, was recorded by the instrument. Using effective area curves for the EPIC pn camera provided in Figure 30 of the XMM-Newton Users Handbook~\citep{ESA:2011} (indicating an effective area up to $\sim 3$ orders of magnitude higher than that of the {\em MESSENGER} XRS), we estimate that AM Her would produce a count rate of no more than 0.1 cts s$^{-1}$ in the XRS. This conclusion is supported by an independent calculation using spectral models based on  {\em GINGA} observations of AM Her discussed by~\citet{Bea:1995}, which we find to give a count rate of 0.08, 0.03 and 0.04 cts s$^{-1}$ in the unfiltered, Al and Mg channels respectively. Hence, although AM Her is sufficiently bright to generate a signal in the XRS, it is unlikely to explain the first component of M1-E1 including the first peak. 

We further note that no other features in any of the flybys are explicable with WGACAT sources.

\begin{figure}[t!]
\begin{center}$
\begin{array}{c}
  \includegraphics[width=3.9in]{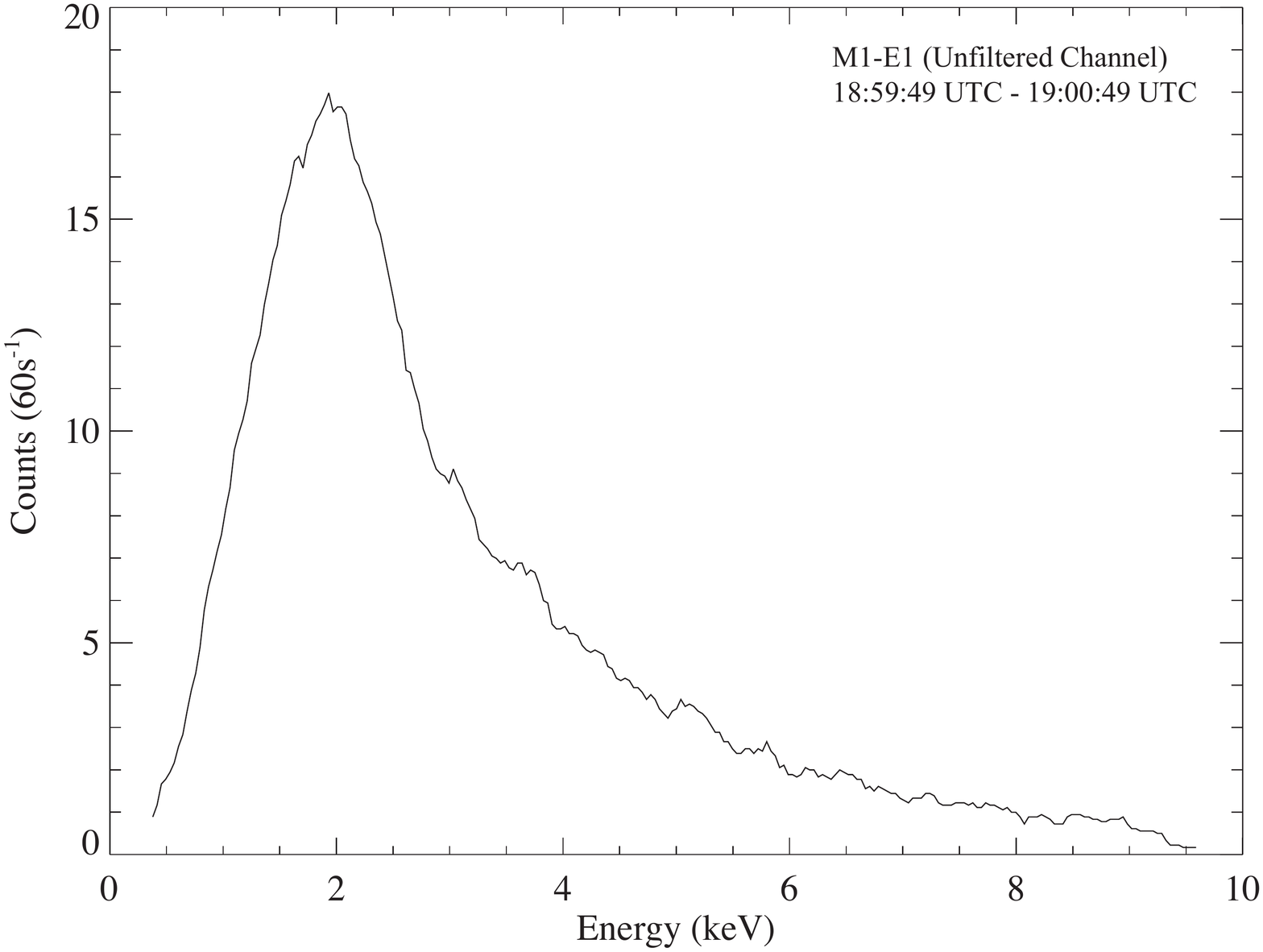}\\
   \includegraphics[width=3.9in]{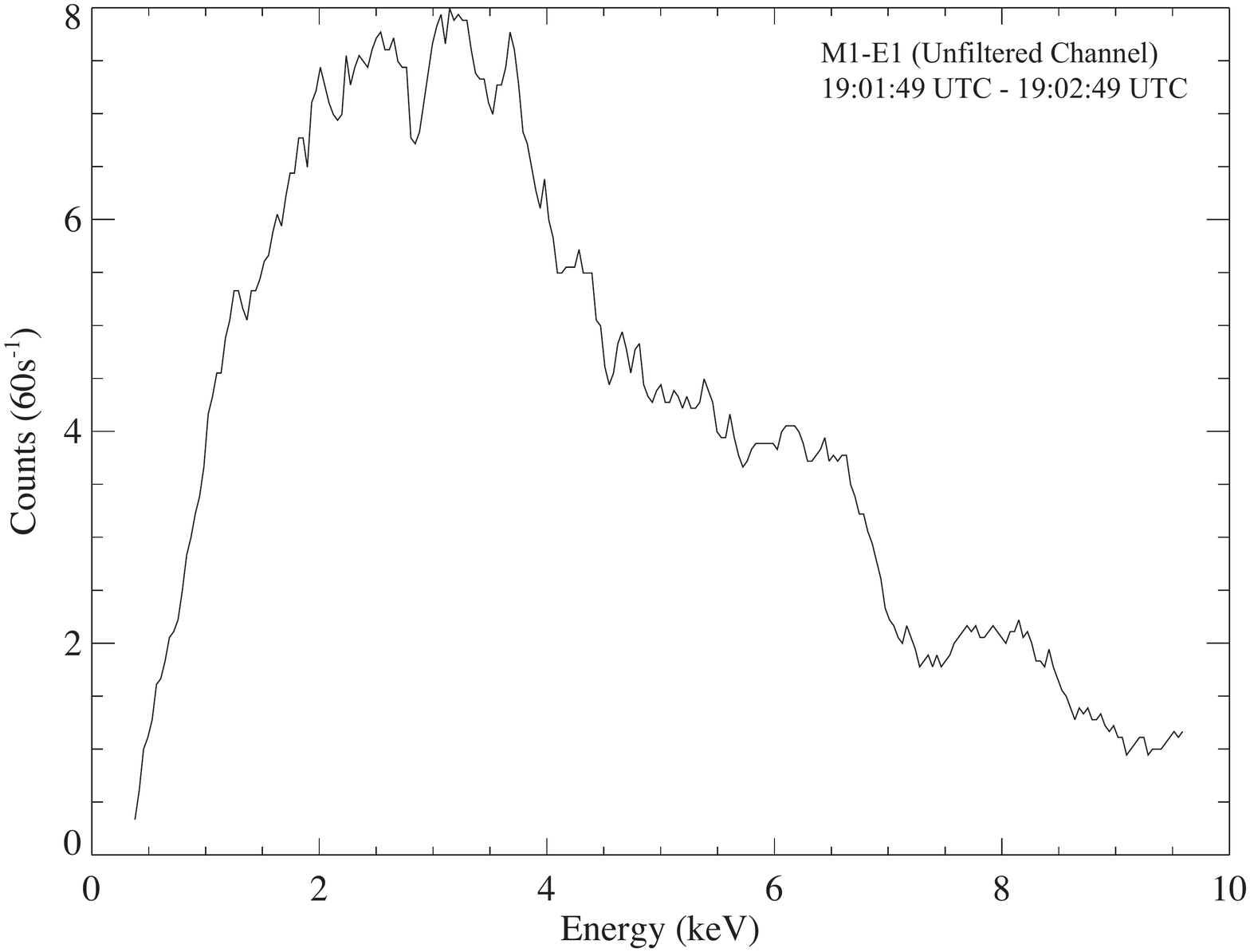}
  \end{array}$
  \end{center}
  \caption{Unfiltered XRS spectra for M1-E1. Top panel: spectrum for the first minute of the event, corresponding to one minute of observations starting at 18:59:49. Bottom panel: spectrum for the final minute in the window, beginning at 19:01:49 UTC. A significant change in the spectral form is seen, corresponding to the transition between events of electron and astrophysical source origin.}
  \label{fig:m1e1spec}
\end{figure}

\begin{figure}[ht!]
\begin{center}$
\begin{array}{cc}
\includegraphics[width=2.6in]{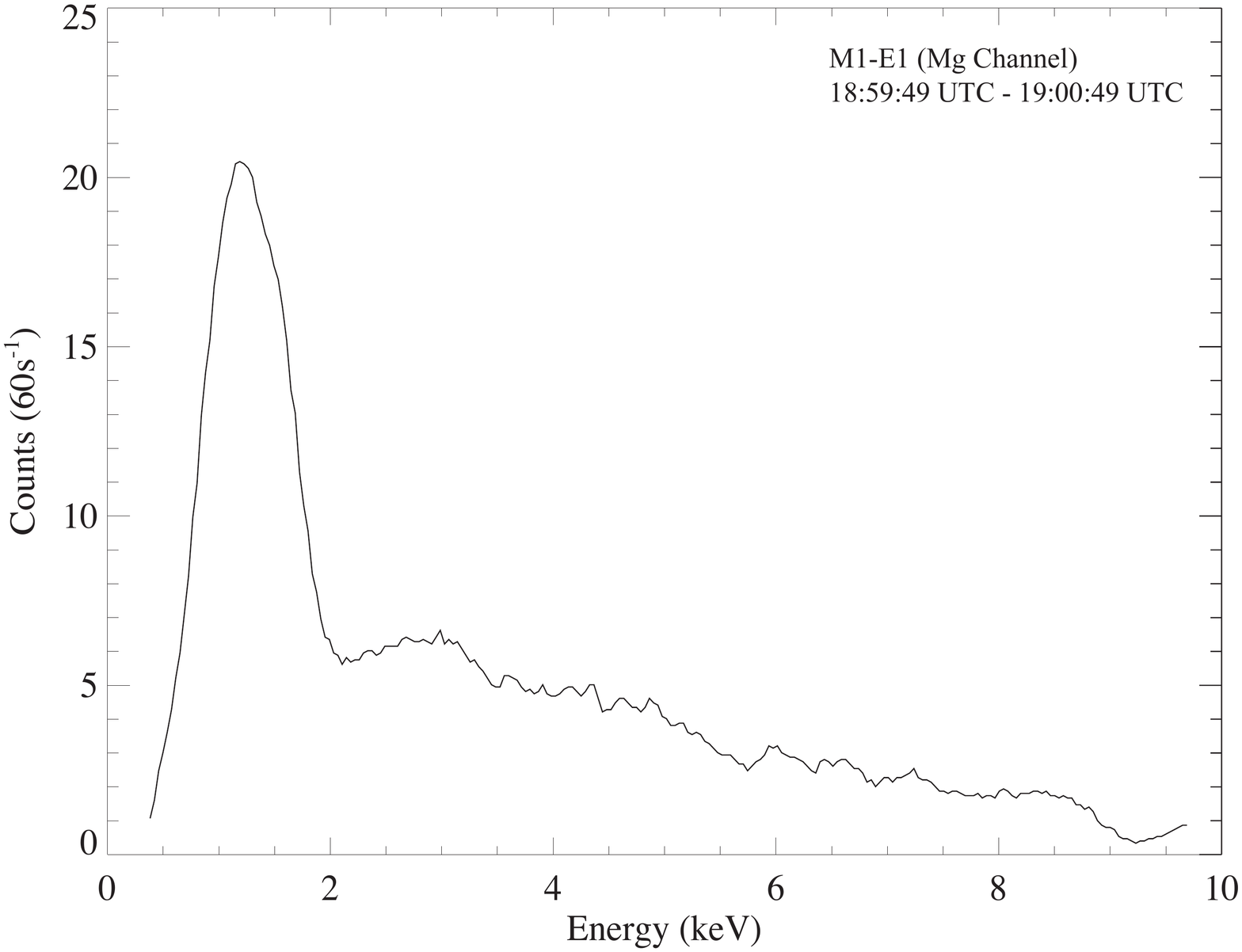} &
\includegraphics[width=2.6in]{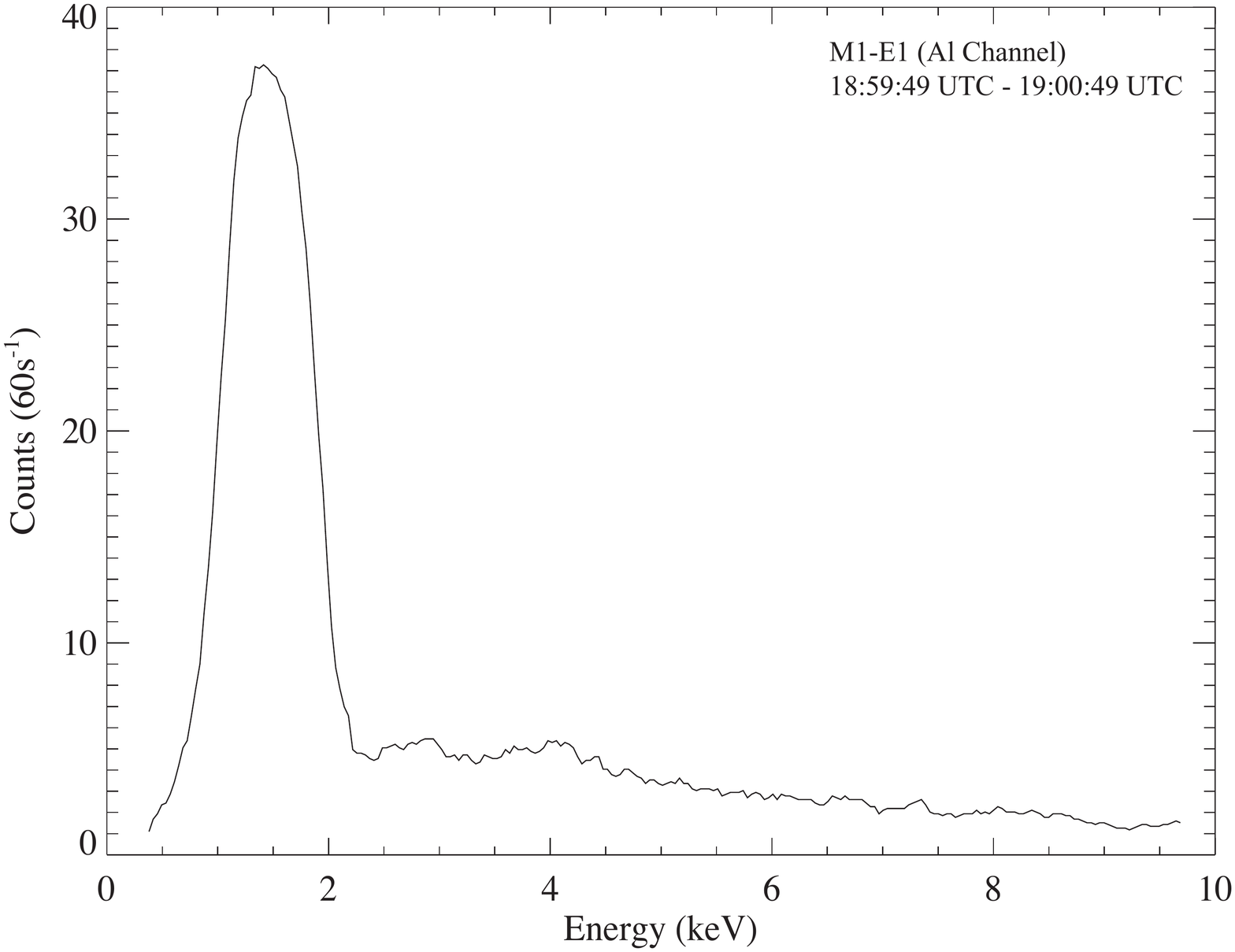} \\ 
\includegraphics[width=2.6in]{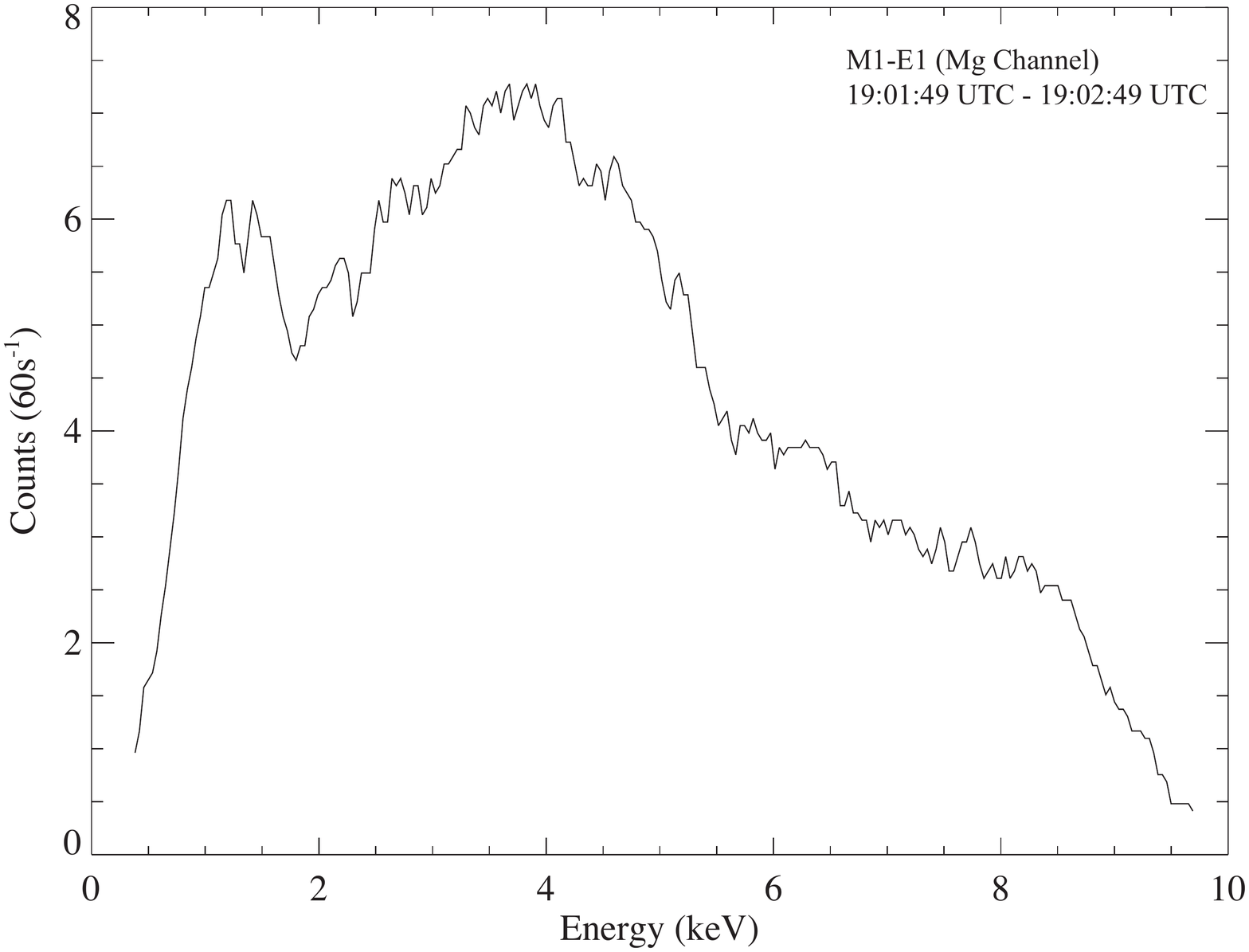} &
\includegraphics[width=2.6in]{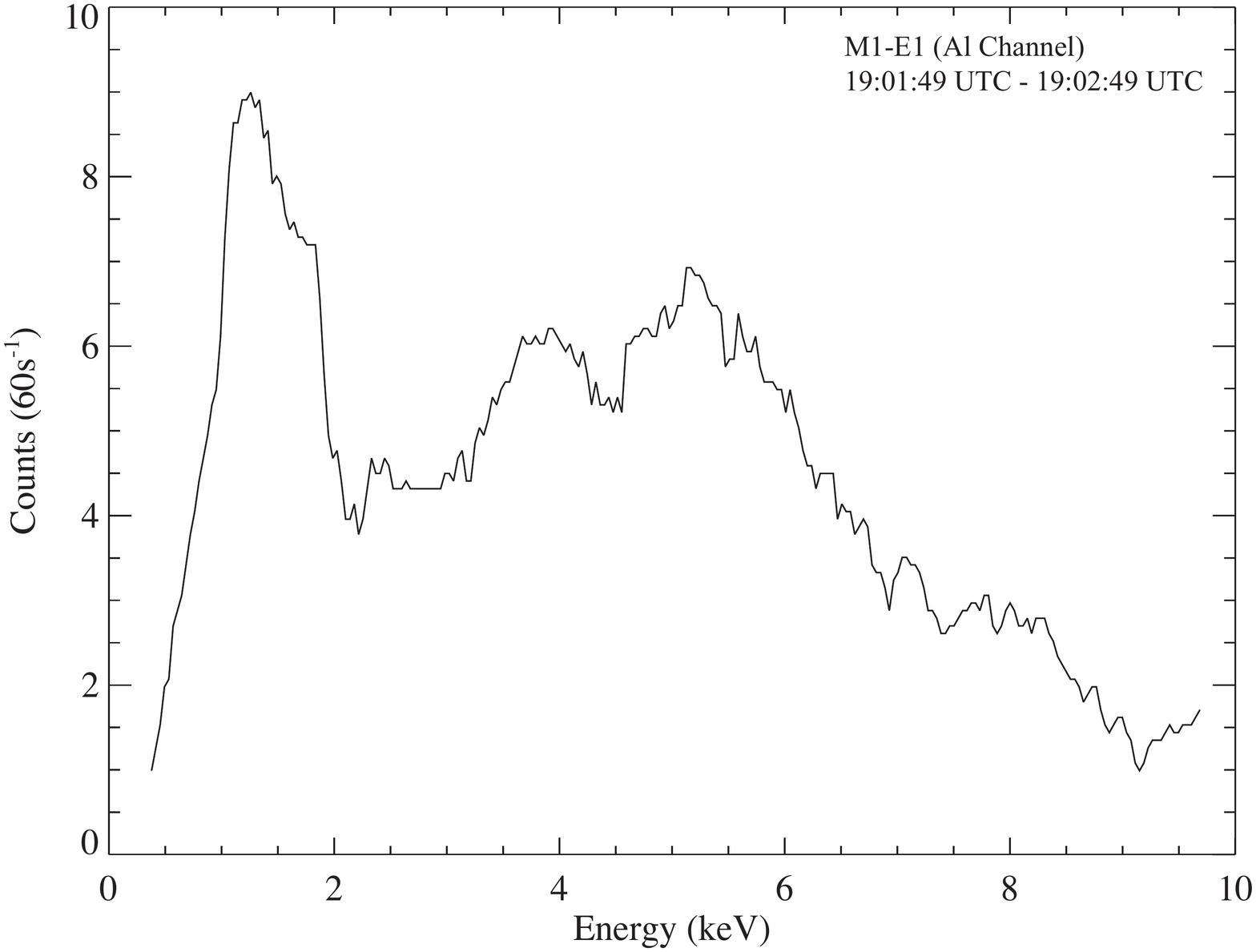}
\end{array}$
\end{center}
\caption{As per Figure~\ref{fig:m1e1spec} but showing data from the filtered XRS channels during the M1-E1 event. Top row: Mg channel (left) and Al channel (right) spectra for the first minute of the event, corresponding to one minute of observations starting at 18:59:49. Bottom row: Mg channel (left) and Al channel (right) spectra for the final minute in the window, beginning at 19:01:49 UTC. A significant change in the spectral form is seen, corresponding to the transition between events of electron and astrophysical source origin.}
\label{fig:m1e1spec_filtered}
\end{figure}

Three spectra are available for the time covered by M1-E1; we show two of these in Figure~\ref{fig:m1e1spec}. The top panel of the figure is the spectrum for the first minute of the event, corresponding to observations starting at 18:59:49 (the beginning of the three minute event window defined by~\citet{Ho:2011}). The spectrum for the second minute is very similar in form, and together these data cover the period which we cannot explain on the basis of astrophysical sources. However, the final minute of the event, beginning at 19:01:49 UTC, shows a different spectral form (lower panel of Figure~\ref{fig:m1e1spec}) with a broad peak near 3 keV consistent with the spectral shape predicted for the LMXB sources described in Section~\ref{sec:m1e1}. This covers the period in which we find the XRS count rate profile well reproduced by the presence of the source GX17+02. We also draw attention to the very significant changes in the spectra recorded by the Al- and Mg-filtered channels of the XRS in the same time windows (Figure~\ref{fig:m1e1spec_filtered}). 

In both channels, the initial minute of the event (Figure~\ref{fig:m1e1spec_filtered}, top row) shows clear evidence of fluorescence peaks from the  K$_{\alpha}$ lines of Mg (1.254 keV) and Al (1.487 keV) as discussed by~\citet{Ho:2011} In the final minute of the event, the spectral form in both channels undergoes a substantial change: the strength of the fluorescence lines is significantly reduced, and there is evidence of a steepening in the spectral slope towards higher energies. 

Thus, the XRS spectra and astrophysical source model are consistent with the suggestion that M1-E1 is an event comprising two phases: an initial count rate enhancement with no astrophysical origin, produced by excitation of sufficient intensity and at an appropriate energy to generate substantial fluorescence in the detector, supporting the electron explanation of~\citet{Ho:2011}, followed by a 20\% - 40\% contribution (based on the Al and Mg channel data in Table~\ref{tab:event_contrib}) in the final minute of the event, comprising photons which do not excite fluorescence in the filters, supporting the astrophysical source argument.

\subsection{M2-E1 and the LMXB GX17+02}\label{sec:m2e1disc}
Our work shows extremely close agreement between the timing of M2-E1 and the presence of three bright sources in the FOV. Two of these, GX5-01 and Sgr X-3, also dominate M2-B1 and reproduce the shape and magnitude of that peak in all three channels. This suggests that our flux estimates for the objects are reasonable, and since they appear in the FOV during the period of M2-E1, a significant fraction of the events in this feature must be astrophysical in origin. However our model does not reproduce the magnitude of M2-E1 well, producing a peak count rate which is approximately one third of the observed value. 

\begin{figure}[ht!]
  \centering
  \includegraphics[width=5in]{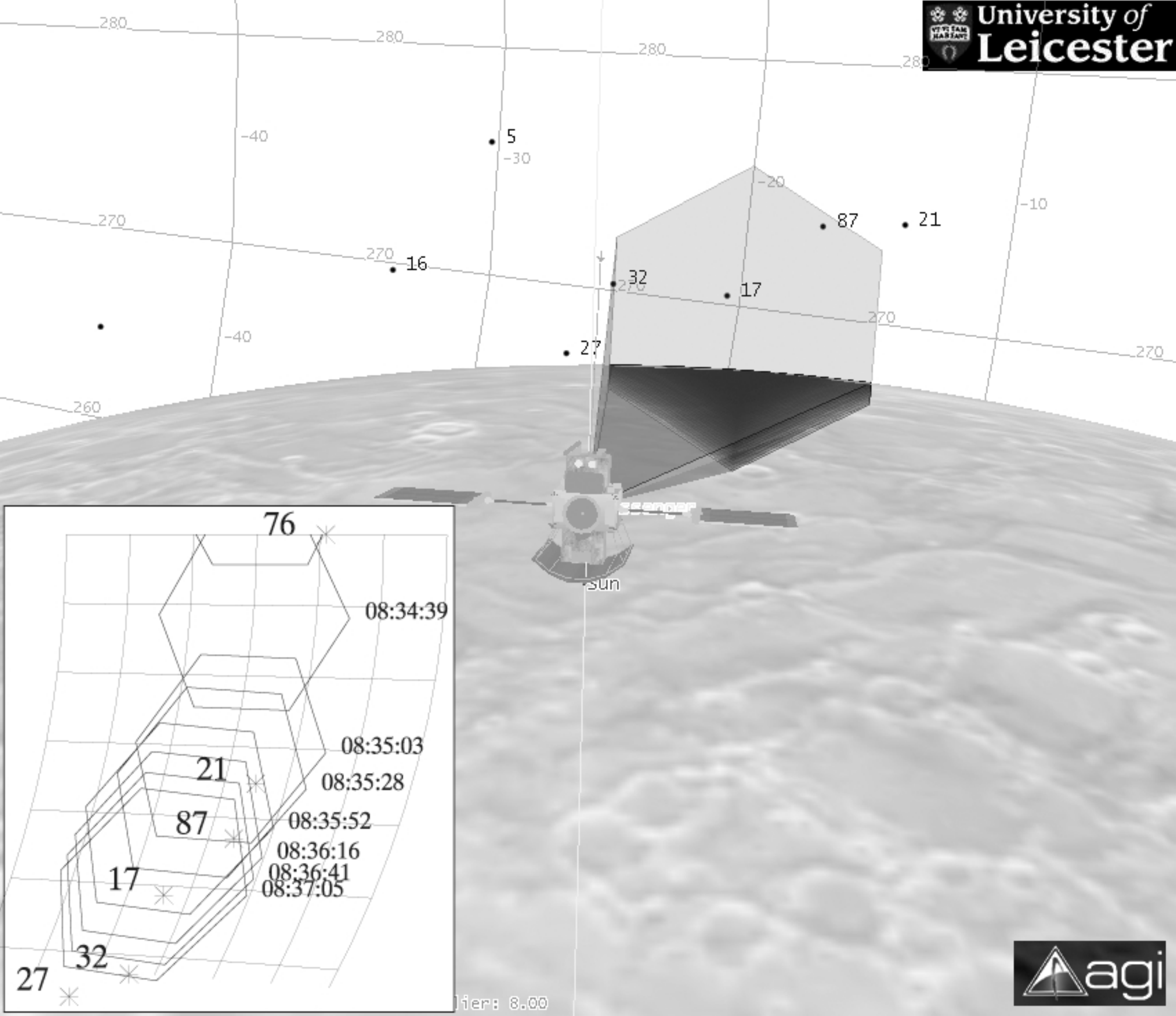}
  \caption{Orientation of the XRS FOV at 08:37:00 UTC on 6th Oct 2008, during the M2-E1 event. This time corresponds to the maximum astrophysical source count rate indicated in our model. Inset: the location of the FOV over time during the M2-E1 event, with times indicated in UTC. Note that the planetary limb is not shown in this figure; the time of first contact between FOV and limb was 08:36:39 UTC. Asterisk symbols indicate the position of bright sources, the numbers corresponding to those given in Table~\ref{tab:sources}. The time of the final FOV position indicated on the plot (08:37:05) is approximately 14 seconds before sources 17 and 32 are occulted by the limb.}
  \label{fig:m2e1stk}
\end{figure}

The third bright source in M2-E1 is GX17+02, which also contributes to M1-E1 and M1-B2 - neither of which shows a substantially under-estimated count rate (the height of M1-B2 in particular is reproduced very well). However, GX17+02 is a neutron star LMXB which exhibits X-ray burst behaviour and is well documented in the literature (e.g.~\citet{Kah:1984,Kuu:2002, Dis:2000,Mig:2007,Far:2005}). If the calculated flux is a good match to features in flyby 1 but not in the later encounter, this may indicate that the object was in different states at these times.

~\citet{Kuu:2002} present a study of GX17+02 and review previous research on this object. Burst durations from a few seconds to 25 minutes have been observed. Our spectral model for GX17+02 was based on the high-state parameters given by~\citet{Chr:1997}, and produces a flux of $2.9 \times 10^{-8}$ erg cm$^{-2}$ s$^{-1}$ in the 1-20 keV band ($2.0 \times 10^{-8}$ erg cm$^{-2}$ s$^{-1}$ in the 1-10 keV {\it MESSENGER} XRS band).~\citet{Kah:1984} report a burst with a peak flux of $9.1 \times 10^{-8}$ erg cm$^{-2}$ s$^{-1}$, lying on top of a steady emission component of $1.56 \times 10^{-8}$ erg cm$^{-2}$ s$^{-1}$ (1-20 keV), observed with the Monitor Proportional Counter instrument onboard the {\em Einstein} observatory. Scaling our GX17+02 spectrum to match the intensity of this burst results in excellent agreement between the predicted and observed peak count rate, and also the time of this peak. However, there are significant differences in the slope of the trailing edge produced by this model compared to the observed profile. Further, the burst flux reported by~\citeauthor{Kah:1984} is much higher than observed for the other bursts in their study, and it is highly improbable that GX17+02 was undergoing a repeat of this extraordinarily large event in the few minutes that the MESSENGER XRS field contained the object. We therefore rule out an intense outburst in GX17+02 as the explanation for M2-E1 in its entirety.

\begin{figure}[ht!]
\centering
  \includegraphics[width=5.1in]{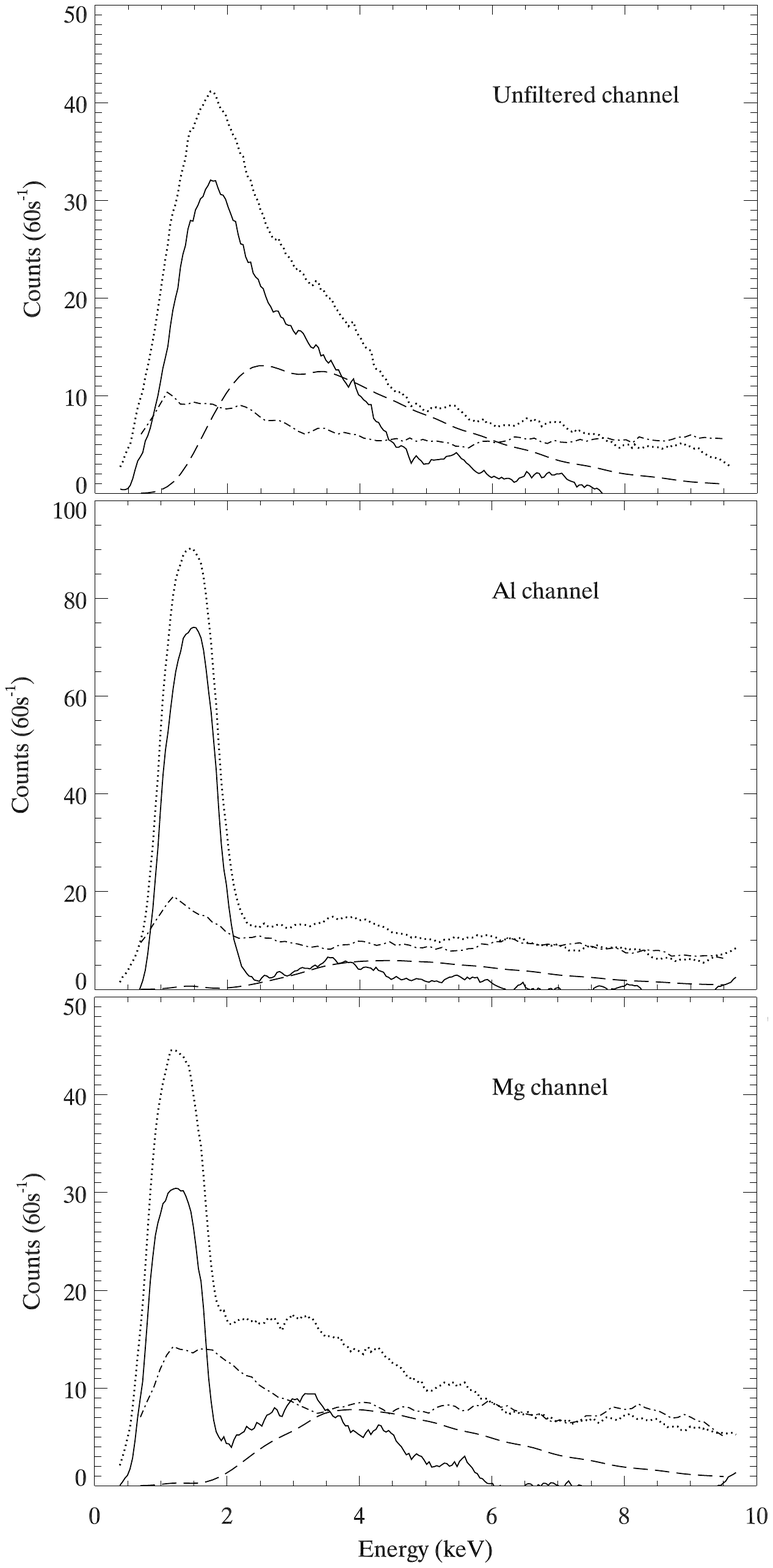}\vspace{-1.8 cm}\caption{Spectra for M2-E1 at 08:35:08 UTC. Dotted histogram: XRS observations. Dot-dash histogram: XRS background level. Solid histogram: XRS observations with background subtracted. Dashed histogram: combined spectrum of the sources in the field of view as calculated in the model.}
\label{fig:m2e1spec}
\end{figure}

The three channel spectra acquired around the peak of the M2-E1 event are shown as the dotted histograms in Figure~\ref{fig:m2e1spec}. For each channel, we have estimated the background level present in the data using XRS measurements from a 60 second interval around 08:27:08 UTC on 6 Oct 2008, a time for which our model predicts no bright sources to be in the field of view (dot-dashed histogram). The resulting background-subtracted observations are shown as the solid histogram. We note that in each channel, the applied background level is marginally higher (1 - 2 counts per bin) than the rates observed at the M2-E1 peak, in some regions above 6 keV; however, the signal levels at these energies are low, and the associated uncertainties such that this feature is not regarded as significant in the context of the current discussion, which is primarily concerned with the general form of the spectrum at energies lower than 6 keV.

LMXB sources are known to exhibit spectral softening in the hard state (e.g.~\citet{Kuu:2002,Mai:2010}). \citet{Mig:2007} consider the spectrum of GX17+02 when in the horizontal branch (HB), normal branch (NB) and flaring branch (FB), each a distinct state of activity with its own X-ray emission characteristics. We have used the HB model given in Table 2 of \citeauthor{Mig:2007} to generate a more detailed, multi-component spectrum for this object, producing a total flux which is similar to that generated by the ~\citeauthor{Chr:1997} model, but in a spectral form which includes a 0.46 keV blackbody emission component that softens the spectral peak by approximately 0.5 keV. The dashed histogram in each of the three channels in Figure~\ref{fig:m2e1spec} shows the profile expected from the combination of sources which pass through the field during this period, weighted by the off-axis angle of each source as a function of time. 

The large fluorescence peaks in the Al and Mg channel observations are absent in the simulations, as is the broader peak around 1.9 keV in the unfiltered channel. Our model does not account for the fraction of X-rays absorbed in the Al and Mg foils which go on to produce fluorescence photons that enter the detector, but these can be approximated on the basis of the photoelectric escape fraction $f$ following the method of~\citet{Ree:1972}. Assuming that the front and rear surfaces of the foil are equivalent, 
\begin{equation}
\label{eq:escape}
f=\frac{1}{2}\omega_K\left(1-\frac{1}{J_K}\right)\left[1-a\ln\left(1+a\right)\right],
\end{equation}
where $\omega_K$ is the K-shell fluorescence yield ($\sim 0.03$ for Mg or Al), $J_K$ is the K-edge jump ratio of the linear absorption coefficient ($\sim 10$ for Mg or Al), and $a$ is the ratio of the mass attenuation coefficients for the incident and fluorescent photon. We find $f\sim 0.01$,  so that approximately 1\% of the flux absorbed in the Al and Mg foils with energies equal to or greater than the K-edge of the material can be expected to generate a fluorescent photon event in the detector. Using the integral model flux between these limits, we find that $\sim 7$ fluorescent photons are expected from the Mg foil, and $\sim 11$ from the Al foil, in a 60 second time integration. Thus, while the astrophysical source spectrum in M2-E1 is expected to generate a fluorescence signal, its strength must be substantially lower than the intense fluorescence peaks which span several bins in the observed spectra. We note also that these spectra provide further evidence against the intense LMXB outburst explanation, which would produce a substantial enhancement of the 3 - 6 keV flux and which is not observed in the XRS spectra. Nevertheless, we highlight some correlations between the background-subtracted XRS observations for M2-E1 and the model spectra: for example, the similarity in the model and observed spectral slopes in the 2-4 keV band of the Al and Mg channels, and the broad general agreement between modelled and observed signal levels beyond 2 keV. We also note that the decay of the unfiltered spectrum above ~4 keV broadly follows the trend predicted by the model. Thus, while there are clear differences between observation and prediction, we find that the comparisons are sufficiently close that the astrophysical explanation for part of M2-E1 remains viable.

The fact that sources implicated in M2-E1 are also found to generate strong features such as M1-B2 and M2-B1 suggests that a significant number of events in M2-E1 must come from astrophysical sources, and as summarised in Table~\ref{tab:event_contrib}, the identified sources contribute between 30\% - 50\% of the total events in this feature depending on the XRS channel being used as the basis of the estimate. But the inability of our model to explain the intensity of the feature, and presence of strong fluorescence peaks in the Al and Mg data, supports the suprathermal electron interpretation of \citet{Ho:2011}, although a modification to the electron model spectrum which they propose should be considered owing to the presence of an astrophysical background signal. It is highly unlikely that the electron event began and ended at precisely the same time that the sources were in the field, but this issue can be resolved by recognising that only $\sim 4$ data points describe this event in the XRS data (as indicated in the top panel of Figure \ref{fig:flyby1}). The start and end times of the event are well matched by our model, but the peak data point (corresponding to the second time bin in the event) and the point one minute later, suggest that a burst of electrons arrived near the end of the period covered by the first time bin, continued through the second time bin, and was decaying in the period of the third. The duration of such an event would be similar to that seen in M1-E1.

\section{Conclusions}
We have used spacecraft telemetry to reconstruct the position of the {\em MESSENGER} XRS field of view as a function of time during the Mercury flybys of 14th January and 6th October 2008, and 29th September 2009. The appearance within the FOV of any of the 100 brightest sources in the {\it ROSAT} bright source catalogue has been recorded. The count rates expected from these sources as a function of time have been calculated, and the results compared to XRS observations from these encounters. In the case of the first two flybys where comprehensive data are available, our model leads to a predicted XRS count rate time series which is in very good general agreement with the observations in all three instrument channels.

Feature M2-E2 cannot be explained by our model since the instrument footprint is entirely located on the planet at this time. The spectrum of this event is different to that observed at other times in the data, and our results support the work of~\citet{Ho:2011} who find that this feature is produced by suprathermal electrons entering the instrument from the Hermean magnetosphere.

Our model shows significant astrophysical contributions to features M1-E1 and M2-E1, also attributed to suprathermal electrons by~\citet{Ho:2011}. In each case we support the general suprathermal electron explanation of \citet{Ho:2011}, and acknowledge that the presence of the energetic electron population discussed in that work is confirmed in research presented by ~\citet{Ho:2011b}.  We suggest, however, that the derived electron fluxes should be reassessed to account for the astrophysical events present in the data. In the case of M1-E1, we find that the first two minutes of the event are characterised by suprathermal electrons, but the final minute is dominated by astrophysical contributions, so the electron flux calculations should be based on the first two minutes of data only.

We find that the onset and end times of M2-E1 are very well reproduced by the appearance of three bright sources in the FOV, and that the end of the event corresponds to the occultation of sources by the planet's limb. Two of the sources contribute to other features of entirely astrophysical origin, lending support to our claim that a significant fraction of the counts in this event are of an astrophysical nature. However, the predicted intensity of the feature is approximately 30\% of the observed value in the Al channel ($\sim50$\% in the Mg channel) and strong fluorescence signals appear in the Al and Mg channels at the peak of the event that cannot be explained by our model. Flaring of the LMXB GX17+02 is ruled out as an explanation, but we propose that the count rate and fluorescence peaks can be explained if the astrophysical photon spectrum is augmented by an excess of suprathermal electron origin, beginning near the end of the first minute that the sources are in the field and has a duration similar to that seen in M1-E1. Therefore in the case of M2-E1, suprathermal electrons still offer a plausible explanation for a substantial part of this feature, but electron flux calculations based on this feature should include a correction for the significant signal contribution received from the astrophysical sources which we have identified. 

Our study highlights two more general points. First, that the analysis of XRS data from {\it MESSENGER}, and by inference other planetary missions with similar instrumentation and observation characteristics (such as the MIXS instrument on {\it BepiColombo}, to which the authors are affiliated), is sensitive to the presence of ``light pollution'' from background sources in the FOV. A detailed consideration of the relationship between the FOV and the position of bright astrophysical sources, as described here, should be included as part of the routine data reduction pipeline whenever instrument fields of view contain background sky.

Finally, the presence of astrophysical sources in XRS data has been presented here as a form of contamination to be accounted for in the analysis of planetary science observations. However, these signals are also ``free'' data obtained as a byproduct of the primary observations, and may contain information on source activity, as well enabling other useful studies such as occultation measurements which can improve knowledge of source locations and terrain profiles. {\em MESSENGER} is not unique in this respect - many other missions generate observational data that include signals which are of little interest to the core mission teams but which would be useful in other areas of planetary science and astronomy. The means of source identification are readily available through resources such as SPICE, and combined with instrument performance data, which in the case of the {\it MESSENGER} mission have been provided via the PDS, these serendipitous observations could form a significant source of data for researchers in different parts of the science community.

\section{Acknowledgements}
The authors thank J. Pye, S. Vaughan, R. Willingale and G. Wynn of the Department of Physics \& Astronomy, University of Leicester, for useful discussions during the preparation of this paper. We are grateful to Richard Starr, MESSENGER XRS Instrument Scientist, for providing information on the performance of the XRS and helpful comments offered on the draft manuscript. We also thank the referees for their constructive comments and suggestions which have improved this work. This research has made use of the Leicester Database and Archive Service (LEDAS) at the Department of Physics and Astronomy, University of Leicester, UK; the Planetary Data System (PDS) Geosciences Node hosted by Washington University in St. Louis, USA; the PDS data archive hosted by the NAIF group at JPL; the SIMBAD database operated at CDS, Strasbourg, France, and NASA's Astrophysics Data System. Finally, we acknowledge with thanks the American Association of Variable Star Observers for the optical data on AM Herculis activity collected by observers worldwide and used in this research.


\bibliographystyle{model2-names}
\bibliography{refs}


\appendix \section{Spacecraft coordinate data sources}\label{sec:appendix}
The MESSENGER spacecraft trajectory and attitude data used in this work were obtained in the form of SPICE ``kernels'' (data files) for the spacecraft, sourced from the Planetary Data System archive (http://naif.jpl.nasa.gov/ naif/data\_archived). Table \ref{tab:kernels} summarises the kernels used in the analysis, including data common to the entire time period (left column) and the three flyby-specific data sets (right column).
 
\begin{table}[ht!]
\centering
\small
\begin{tabular}{l|l}
\hline

Common & Flyby-Specific \\ \hline
lsk/naif0009.tls 	&  ck/msgr\_0801\_v02.bc \\
pck/pck00008.tpc & (Flyby 1)  \\
pck/pck00008\_msgr.tpc \\ 
sclk/messenger\_0766.tsc & ck/msgr\_0810\_v01.bc\\ 
fk/msgr\_dyn\_v600.tf & (Flyby 2)\\
fk/msgr\_v200.tf \\
ik/msgr\_xrs\_v001.ti & ck/msgr\_0909\_v01.bc \\  
spk/de405.bsp  & (Flyby 3)\\
spk/msgr\_040803\_091031\_120401.bsp \\
\hline
\end{tabular} 
  \caption{SPICE kernels used to determine {\it MESSENGER} position and attitude. Left column: kernels common to the analysis of all three flybys. Right column: flyby-specific kernels providing spacecraft attitude information for individual encounters.}\label{tab:kernels}
\end{table}

\end{document}